\documentstyle[amssymb,amsmath]{article}
\begin{document}

\def\slash#1{\setbox0=
\hbox{$#1$}#1\hskip-\wd0\hbox to\wd0{\hss\sl/\/\hss}}

\centerline{\Large\bf
Geometric Bounds on Kaluza-Klein Masses 
}
\vskip 1truecm

\centerline{
Avijit Mukherjee\footnote{E-mail: avijit@sissa.it}
{\it and} \ Rula Tabbash\footnote{E-mail: rula@sissa.it}
}
\bigskip
\centerline{\it
Scuola Internazionale Superiore di Studi
Avanzati (S.I.S.S.A.)}\centerline{\it
Via Beirut $2$-$4$, I-$34014$ Trieste, Italy$^{1,2}$}
\bigskip
\centerline{\it
Istituto Nazionale di Fisica Nucleare, 
Sez. di Trieste}\centerline{\it
Via Valerio $2$,
I-$34127$ Trieste, Italy$^2$}
\vskip 1.2truecm
\begin{abstract}

\noindent
We point out geometric upper and lower bounds on 
the masses of bosonic 
and fermionic Kaluza-Klein excitations in the context of theories with 
large extra dimensions. 
The characteristic compactification length scale
is set by the diameter of the internal manifold. Based on
geometrical and topological considerations, we find that 
certain choices of compactification manifolds are 
more favoured for phenomenological purposes.
\end{abstract}
\vskip 1truecm

\section{Introduction and Summary}
In the recent past, there has been much attention to models with 
large extra dimensions. The surge in activity surrounding this 
idea owes its origin to the belief
that the existence
of extra dimensions (beyond four) seems to be a crucial 
ingredient for the unification of gravity with gauge forces.  
The initial goal of large radius, $r\gg M^{-1}_P$, 
compactification schemes is to 
weaken the hierarchy between the electroweak scale, 
and the four dimensional gravity scale, $M_P$. 
The idea is that the 
matter content of the Standard Model of elementary particles (SM)
is confined to $(3+1)$-dimensions, as suggested by
\cite{rub,dvali,A1,akama}, while gravity lives in the whole 
$D$-dimensional space ($D>4$). Upon compactification, 
the hierarchy problem 
is solved by lowering the 
fundamental scale of gravity, $M_*$, down to TeV through
a model-dependent relation between $M_*$ and $M_P$. 
The compactification mechanisms suggested so
far, can be classified into two broad categories: models with 
tensor product of our four dimensional world with the internal space
\cite{dvali,A1} (in line with the original Kaluza-Klein
ideology), and models with warp product \cite{wett,rand} between the 
same.\\

\noindent
On compactifying down to four dimensions, one may 
in general get new degrees of freedom added to the 
SM
spectrum. The new states can be purely from the gravitational
sector, or have Standard Model KK excitations in addition 
(depending
on
whether the SM interactions are written directly in four dimensions,
using the induced metric, or written fully in $D$ dimensions).
In any case, the new states might 
lead to detectable modifications of 
the existing accelerator data and cosmological observations
\cite{dvali2}. The phenomenology of both categories has been 
investigated
in great pursuit during the past two years, 
and we refer to a summary of the recent findings in \cite{hewett}.
The usual way to avoid such new contributions 
to the prevailing scenario at low energies 
is often either by decoupling the particles 
by making their masses very heavy (beyond 
the present reach of accelerators, say $\gtrsim$ TeV),  or 
by imposing 
judicious bounds on their couplings and masses. 
Recently, it was suggested \cite{kn:m} 
that the heavy masses could be realized
naturally (without fine-tuning), 
utilizing only certain geometrical 
properties of the internal manifold, namely 
that the masses arising from 
compactification are exponentially large being related to the 
volume of the internal hyperbolic manifold.\\

\noindent
In this work we scrutinize the
criteria for chosing the internal manifold, in both cases 
of tensor and warp product compactifications, based on 
geometrical and topological 
arguments,
such that the unwanted KK contributions 
are avoided.
We focus on compact,
connected, and smooth internal manifolds with scalar curvature,
bounded from below $\kappa\geq (d-1)K$, where $K$ is a constant. 
We consider, for generality, a gravity theory coupled to a Dirac
spinor in the presence of a gauge theory. 
This consideration, 
has a final aim to be applied to the SM.
However, in order to keep the discussion
simple and sufficiently general (model independent) we shall
not concern ourselves with finer details like 
localization mechanisms, the issue of obtaining chiral 
fermions starting from odd dimensions,...{\it etc}. 
Instead, after performing the general 
analysis of various bounds on the KK masses, 
one may specialize to the case of the SM.\\

\noindent
Most of our analysis relies on the following basic observations 
concerning 
the spectrum of Riemannian manifolds, and the Dirac operator 
on spin manifolds. The main fact is that the spectrum 
of $\slash{D}$ and that of the Laplacian on compact Riemmanian manifolds 
is discrete, bounded from below, and the eigenvalues (counted with 
multiplicity) are ordered: 
$0=\lambda_0\leq\lambda_n\leq \lambda_{n+1}$. 
Moreover, there exist lower bounds on 
$\lambda_1$ of a Laplacian acting on a scalar in the compact 
manifold. In addition, there are 
upper bounds on the eigenvalues which sets a ceiling to how
heavy they can be lifted.
These translate into lower bounds on 
the 4-dimensional 
tree-level
masses of particles arising from compactification.
For spinors, 
the classic theorem of Lichnerowicz enables us to impose 
similar bounds, upper and lower, and altogether
exclude tree level massless fermions
for certain internal manifolds. To sum up, 
we use topological considerations
to comment on bosonic KK zero modes, while we use geometrical 
arguments to impose bounds on fermions and massive KK modes.\\

\noindent
The paper is organized as follows: in section 2, we  
summarize our conventions, and state our requirements for 
choosing the internal manifold so that we are able to produce a 
phenomenologically reliable scenario in a rather model 
independent way.
The implementation of these demands is carried on the successive 
sections. 
In section 3, we relate the eigenvalues of the Laplacian on the
internal space with the tree-level masses in four dimensions,
and discuss geometrical upper and lower bounds for massive bosons,
and topological conditions for  
the massless ones. In section 4, we comment on massless spinors 
using Lichnerowicz theorem, 
and point out the existence of curvature-dependent 
upper and lower bounds on massive ones.
In order to satisfy all of our requirements, stated in the second section,
we are able to rule out certain choices of the
compactification manifolds. Finally, we summarize our conclusions
in section 5.

\section{Conventions and Set up}
As mentioned in the introduction, 
we consider Einstein's gravity coupled to a Dirac
spinor and a Yang-Mills gauge theory
on
a $D$-dimensional 
manifold $W=M_4\cup Y$, 
$$
S=\int\limits_W d^D x\;\sqrt{-g}
\left[\frac{1}{G}{\cal R}+\frac{1}{4}F^2+i
{\bar \psi}\hat{\slash{D}}_A\psi\right]
$$
where $M_4$ is a four manifold, which we eventually 
identify as our 4-dimensional world, and $Y$ is a compact
$(D-4)$-dimensional manifold.
$\hat{\slash{D}}_A$ is the twisted Dirac operator on $W$ 
and $F$ is the YM
field strength
(For a detailed treatment
of an analogous setting in six and ten dimensions, 
and conventions,
we refer to \cite{rss}).
Consequently, the various
fields on $W$ will be decomposed as the following:
scalars on $W$ will be scalars on both $M_4$ and $Y$;
a vector on $W$ will be a vector on $M_4$ and scalar on 
$Y$ or vice versa; the graviton on $W$ will
appear as a graviton on $M_4$ and scalar on $Y$ or vice
versa, or as a vector on both submanifolds. Finally,  
a spinor on the parent manifold will decompose as a spinor 
on both $M_4$ and $Y$. It is perhaps worth mentioning that  
a spinor defined on $W$ which is a fibre product of 
$M_4$ and $Y$, does not necessarily split into spinors 
defined on the two submanifolds individually, as it does 
in the Cartesian
tensor product case. However, in the special case 
of a warp product, the 
fibration being trivial, this decomposition once again holds.\\

\noindent
Our analysis includes both the tensor product decomposition
(the usual compactification case), 
and the warp product decomposition \cite{lawson}.
Let us recall that for the tensor product, $W=M_4\otimes Y$, 
the inherited metric is $
{\hat g }= g_4 + g_Y
$ where $g_4$ and $g_Y$ are the metrics on $M_4$ and $Y$ respectively.
Whereas for a warp product, $W=M_4\otimes_{\Bbb R^+} Y$,
the resultant 
inherited metric is of the form ${\hat g}=f^2 g_4+g_Y$, 
where $f$ is a smooth map $f:Y\rightarrow {\Bbb R^+}$.
In this work we choose the warp factor to be 
$f= e^{-\frac{1}{2}\phi}$ as in \cite{rand}. The warp factor is to 
be consistently determined by solving Einstein's equations.\\

\noindent
Our main requirements, for the theory resulting 
after compactification,
can be outlined as the following:
\begin{description}
\item[($i$)]{With respect to the gravity sector, 
we want to
end up with one massless graviton, no additional massless
gauge bosons, and no massless scalars\footnote{Massless 
vectors may
enhance the gauge symmetry, and gravitational interactions
mediated by scalars violate the equivalence principle.}.} 
\item[($ii$)]{The zeroeth KK mode(s) of the Dirac 
spinor is massless in four dimensions. 
If specialized to the SM,  
this translates into the requirement that
fermion masses are 
exclusively due to Higgs 
mechanism.}
\item[($iii$)]{The masses of KK excitations of various fields 
are naturally heavy.}
\end{description}

\section{Bounds on bosonic KK masses}
The starting point of our analysis is to examine the masses in 
the gravity sector. 
Looking at the linearized Einstein's equations on $W$\footnote{
We use the linearized equations for obvious reasons, nevertheless
we include for completeness
the decomposition of $\cal R$ on $M_4$ and $Y$,
$${\cal R}={\cal R}_4+\kappa$$
for tensor product; and for warp product \cite{lawson}
$$
{\cal R}=\frac{1}{f^2}\left\{ {\cal R}_4-8f\Delta f-6\parallel
\nabla f\parallel^2  
\right\}+  \kappa 
$$
where $f$ is the warp factor, which we may assume to 
be $f=e^{-\frac{1}{2}\phi}$.}, 
${\hat \Delta}h_{{\bar k}{\bar l}}=T_{{\bar k}{\bar l}}$  ($\bar k$'s are 
the indices
on $W$),  
one can relate the spectrum 
of $\Delta_Y$ with the 
tree-level masses of the various fields on $M_4$.
The parent 
Laplacian\footnote{The ``hat'' superscript refers to quantities defined on 
$W$.}, $\hat{\Delta}$, decomposes as
\begin{equation}
\hat{\Delta}={\Delta}_4  +{\Delta}_Y
\label{eq:g}
\end{equation}
in the tensor product case, and as 
\begin{equation}
\hat{\Delta}= e^{\phi(y)}\Delta_4 +\Delta_Y-\frac{1}{2}\left(
\partial_l \phi(y)\right)\partial^l
\label{eq:l}
\end{equation}
in the warp product case ($l$'s are the indices on $Y$).

\subsection{Massless bosons}
A necessary condition
to meet the first demand ($i$) is to select $Y$ with the appropriate
betti numbers\footnote{
The number of zero modes of the Laplacian 
(or equivalently the dimension of the space  harmonic $p$-forms) 
on a compact manifold $Y$ are 
given by the $p$-th betti 
numbers $b_p(Y)$ of the manifold.}. 
Since $b_0(Y) = 1$ for any general connected manifold $Y$,
we are guaranteed to end up with one massless 
graviton on $M_4$. $b_1(Y)=0$ would ensure that no new massless
vector bosons, as well as massless scalars
are produced in $M_4$ after compactification. 
However, this is not the case for 
a general $Y$. For example, a circle has $b_1 (S^1)=1$ and 
for a torus, 
$T^d$; $b_1=d$, both of which therefore admit massless 1-forms. 
$S^d$s are in general suitable ambient spaces for performing such 
compactifications, since the have $b_1(Y)=0$ for $d>1$.
Other possible alternatives are Calabi-Yau's, $K3$'s, 
suitable orbifolds
of the type $T^d/Z_n$,
compact hyperbolic manifolds for $d\geq 3$. 
In general, for spaces having $b_1\neq 0$ 
quotienting by an appropriate discrete isometry, 
often leaves us 
with $b_1=0$. This topological classification
is insufficient when harmonic spinors are discussed to 
meet the demand ($ii$). For, in that case, the 
curvature of the manifold (a geometric parameter) plays the 
decisive role.\\

\noindent
This analysis of the massless sector  
applies to both compactification schemes-- tensor or warp.

\subsection{Massive bosons: Lower bounds}
Having considered the massless fields, which are the zero modes of 
the Laplacian on the internal space, we now turn our attention to 
the first massive excitations.  
As we mentioned in the introduction, there has been an extensive
study for the first non-zero eigenvalue of the Laplacian on Riemannian 
manifolds. Rigorous bounds, particularly 
for manifolds  
with scalar curvature bounded from below by $(d-1)K$  
(where $K$ is constant and $d$ is the dimension of the manifold),
have been established.
Any way, assuming a slowly varying $\kappa$ makes it possible
to replace it by $(d-1)K$ in the context of discussing mass bounds
and scales.
It may be noted that, the eigenspectrum being strictly ordered,
only the lowest massive states are relevant to 
our analysis, because if we achieve to decouple these, 
then all the 
higher modes will be automatically eliminated in the effective four 
dimensional theory.\\

\noindent
Let $Y$ be a compact manifold, and $\lambda_1$ the first
non-zero eigenvalue of $\Delta_Y \phi_n=\lambda_n \phi_n$, where 
$\phi_n$ is a scalar. Then
\cite{yau} 
\begin{equation}
\lambda_1 + \mbox{max} \{  -(d-1)K, 0
\} \geq \frac{\pi^2}{4\sigma^2}
\label{eq:w}
\end{equation}
where $\sigma$ is the diameter of the manifold. 
It is worthwhile to note from (\ref{eq:w})  
that the fundamental parameter for masses arising from compactification
is $\sigma$, and not generically the 
volume of the manifold, as it is commonly 
thought\footnote{This can be easily
understood by observing that it is possible to change the spectrum of
the Laplacian by deforming the manifold, and yet keeping its 
volume 
fixed. However, the relation between $M_*$ and $M_{P}$ will 
always involve the volume of $Y$.}. However, in certain cases 
one can proceed a step further
and relate $\sigma$ to the volume of the manifold, and hence 
rewrite 
the bounds in terms of the volume instead (e.g. in $S^d$ and certain
compact 
hyperbolic manifolds).
The inequality (\ref{eq:w}) translates
effectively into a statement about the bounds on the 4-dimensional
masses\footnote{Here, 
and elsewhere, we use the rest frame when referring to massive states.} 
of the lowest
excitations, $m_1^2$.
It is obvious
that when the Ricci curvature, $\kappa$, of $Y$ is non-negative, 
then one recovers
the standard scenario:
$\lambda_1\geq \pi^2/(4\sigma^2)$,
where in the standard KK scenario, as in \cite{dvali}, 
$\sigma$ is identified with the diameter
of the compactification circle(s).\\

\noindent
At this point, we note that
the explicit expression of the bounds will depend on 
the nature of the product between the two manifolds-- 
a tensor or a warp 
product. In the tensor product case, the bound (\ref{eq:w}) on 
first scalar excitations (in the 4-dimensional effective theory) 
will 
remain unaltered,
$$
m^2_1\geq \frac{\pi^2}{4\sigma^2}-\mbox{max} \{  -(d-1)K, 0
\}
$$
A natural choice would be 
$\sigma^{-1}\sim M_*$ (say $\sim{\cal O}(\mbox{TeV})$).
In the case of $\kappa$ bounded from below by a negative
constant (i.e. not everywhere positive), the bound will involve
the infimum of the curvature (or the curvature itself, if
constant or slowly varying), and in order to achieve $m^2_1
\gtrsim \mbox{TeV}^2$ we need
\begin{equation}
\kappa\approx |(d-1)K|\lesssim (\frac{\pi^2}{4}-1)\mbox{TeV}^2
\label{kappa}
\end{equation}    
It is remarkable that satisfying this bound on the curvature 
requires no fine-tuning at all\footnote{For large extra dimensions
models to do anything with string/M-theory, one must have
$d\leq 7$.}.
\noindent
As was noticed in \cite{kn:m}, some manifolds with negative scalar
curvature, like compact hyperbolic ones, may have attractive features    
like exponentially large KK masses.       
We would like to speculate that negatively curved internal 
spaces may also be favoured (beside the  
string inspired Ricci flat compactifications), 
because they support the existence of massless spinors, 
as will be shown in the next section.\\

\noindent
In the
warp product case,
it is not straightforward 
to comment at this level. Mainly, 
because the eigenvalue of $\Delta_4$, $\tilde m_1^2$, 
will be $y$-dependent
in the $D$-dimensional theory,
\begin{equation}
{\tilde m_1^2}\geq e^{-\phi(y)} 
\left[
\frac{\pi^2}{4\sigma^2}-\mbox{max} \{  -(d-1)K, 0
\}
-\frac{1}{2}\left(
\partial_l \phi(y)\right)\partial^l\right]
\label{l0}
\end{equation}
and  
the integration over the $y$ coordinates, in order to 
get the effective four dimensional mass, becomes non-trivial
in the
the presence of the extra terms.
It is clear from (\ref{eq:l}) that both the warp factor and 
the term 
$\partial_l \phi(y)\partial^l$ (which should be understood as 
the gradient of the wave function in the internal space) 
will change the interpretation  
of the effective four dimensional mass, and hence it is not 
straightforward 
to make a statment about the bound.
In fact, it is not only the warp factor dependence on $y$ which 
matters here, 
but also the gradient of the wave function 
of the field 
in the internal space. 
This conclusion is in contrast with the bounds on graviton 
excitations
discussed in \cite{kn:m}.\\

\noindent
Whereas the third demand, of the scalar 
sector in the theory\footnote{This bound applies on {\it any}
massless 
scalar field in the theory, whether or not in the gravity 
sector.},  
can be met in the tensor product case 
(by chosing $Y$ with an appropriate
diameter)
it seems difficult 
to be fulfilled without further model dependent details,  
in the warp product case.\\

\noindent
The same arguments can be carried over to the case of 
vectors and rank two tensors arising from gravity sector, 
as a consequence of the linearization procedure. In principle, 
one could expect curvature dependent additions to the equations
of motion, as can be read off from (\ref{lapl}), but 
those additional terms are dropped off because they involve
${\cal{O}}(h^2)$. In this context, we would like to point out
that the vector degrees of freedom, resulting 
from the metric decomposition, can not in general be
eliminated by a gauge choice,
and their coupling to a typical scattering amplitude
are comparable to those of graviton exchange \cite{ss}.
Hence, it is important to make them very massive and weakly
coupled. On the other hand, the curvature dependent terms in (\ref{lapl})
will appear in the YM equations of motion,
and it will be 
difficult to draw conclusions\footnote{
Unfortunately, no rigorous bounds 
for Laplacians on 1-forms, relevant to our discussion, 
have been worked out.}, apart from the special case when 
$d=2$ where the bounds for the 1-forms are the same as in 
the case of scalars \cite{junya}. 
In any case, from (\ref{lapl}), it can be argued
that the bounds
for these fields are of the same order as in (\ref{eq:w}).\\
\label{r}
\subsection{Massive bosons: Upper bounds}
Finally, we would like to add the following remark.
Although the first non-zero eigenvalue of $\Delta$ is bounded
from below, it is not possible in general to push it to an infinitely
heavy scale. There exists an {\it upper} bound which depends on the 
same
parameter $\sigma$. For example, if the
Ricci curvature $\geq 0$ then the $j$th eigenvalue
is bounded from above by \cite{cheng}
\begin{equation}
\lambda_j\leq \frac{2j^2}{\sigma^2}d(d+4)
\label{l1}
\end{equation}
And in the case when Ricci curvature is bounded from below by a 
negative number ($K<0$), then the upper bound \cite{cheng} becomes 
more complicated, and includes the (lower bound of) the curvature. 
For $d\geq 2$
the bound is
\begin{equation}
\lambda_j\leq \frac{(2l-1)^2}{4}K+\frac{4\pi^2j^2}{\sigma^2} 
(1+2^{l-1})^2
\label{l2}
\end{equation}
for $d=2l$, $l=1,2,...$, and
\begin{equation}
\lambda_j\leq l^2K+\frac{4(1+\pi^2)j^2}{\sigma^2} 
(1+2^{2l-2})^2
\label{l3}
\end{equation}
for $d=2l+1$, $l=1,2,...$.
Concerning 1-forms, the same upper bound holds for $\lambda_1^{(1)}$, 
because 
$\lambda_1^{(1)}\leq \lambda_1$ \cite{junya} (without any 
further assumption 
concerning the curvature).\\

\noindent
In the tensor product case, the bounds (\ref{l1},\ref{l2},\ref{l3}), 
remain unaltered 
on the effective four dimensional masses. However, a more careful 
treatment, as discussed in section (3.2),
is required in the warp product case.


\section{Bounds on fermionic KK masses}
The Dirac operator on the spin manifold $W$ acting on spinors is 
$
\hat{\slash{D}}=\gamma^{\bar k}(X)\;\partial_{\bar k}$,
where $\gamma^{\bar k}(X)$'s are the $D$-dimensional 
Gamma matrices
in curved space written in terms of the Veilbeins on $W$.
It decomposes in the tensor product case as
\begin{equation}
\hat{\slash{D}}=\slash{D}_4+{\slash{D}}_Y
\label{eq:c}
\end{equation}
where ${\slash{D}}_Y=\gamma^l(y)\partial_l$, 
${\slash{D}}_4=\gamma^\mu(x)\partial_\mu$; 
$\gamma^l(y)$, and $\gamma^{\mu}(x)$ 
being the Gamma matricies of the 4 and $(D-4)$ dimensional
spaces respectively ($\mu$'s run over $M_4$). In the warp product case, 
the $\slash{D}$ splits differently from the one in (\ref{eq:c}), 
and has the form
\begin{equation}
\hat{{\slash{D}}}=
e^{\frac{1}{2}\phi(y)} {\slash{D}}_4+{\slash{D}}_Y
\label{eq:d}
\end{equation}
where ${\slash{D}}_4$ and ${\slash{D}}_Y$ 
are the same operators defined
previously.
Hence, the four dimensional fermion masses are related to the eigenvalues
of the Dirac operator on the internal manifold\footnote{We define
of the spinor (mass$)^2$ as an eigenvalue of the squared Dirac operator.}.  
And in particular, the observed massless fermions in four dimensions
are nothing but the zero modes of $\slash{D}_Y$ (which lie in $\mbox{ker}
\slash{D}_Y$). It has been shown by Lichnerowicz \cite{L} 
that not all manifolds
admit harmonic (massless) spinors. The argument is based on the relation
between the squared Dirac operator and the scalar curvature,
\begin{equation}
{\slash{D}}_Y^2= 
{\nabla}^*\nabla +\frac{1}{4}\;{\kappa}
\label{square}
\end{equation}
where $\kappa$ is the 
scalar curvature of $Y$, ${\nabla}^*$ 
is the 
adjoint of $\nabla$,
and ${\nabla}^*{\nabla}$
is the connection Laplacian (a positive operator).
We use this theorem not only to identify candidate manifolds in which 
the demand ($ii$) can be realized, but also to set geometrical bounds so 
that ($iii$) is met. 

\subsection{Massless fermions}
Recall that 
a spinor $\psi$ is said to be {\it harmonic} iff $\slash{D}\psi=0$, i.e 
$\psi\in \mbox{ker}\;\slash{D}$, 
It is helpful to remember that   
$\mbox{ker}\;\slash{D}=\mbox{ker}\;\slash{D}^2$, 
and that this space is finite dimensional,
\cite{lawson} and this space is identified
with our space of massless fermions, as metioned above.
It has been shown in \cite{L} 
that the existence of harmonic spinors depend strongly 
on the
scalar curvature of the manifold,  
and in particular, massless spinors 
do not exist on manifolds with a positive scalar 
curvature\footnote{As 
an example, massless spinors do not exist on a sphere.}.
This no-go theorem applies also to cases where the 
scalar curvature 
is non-negative everywhere, and not necessarily constant.
In addition, the formula (\ref{square}) shows that fermion mass square
is bounded from below by the curvature since the operator
${\nabla}^*
{\nabla}$ is positive.\\

\noindent
According to the above argument, meeting the second demand, namely
supporting massless spinors (which will eventually acquire mass
only through Higgs mechanism), rules 
out the entire class of manifolds with positive curvature, unless
they have further discrete isometries.
If one 
wants to relax the second demand, by having the above mass term, 
then 
a careful attention should be paid in order not to spoil gauge invariance. 
For instance, 
a direct mass
term in the action for the SM fermions is not gauge invariant. So, 
adding such a tree-level mass term by hand, and yet keeping gauge
invariance, will be at the cost of doubling (or increasing) 
the number of degrees of freedom (and hence the number of SM generation) 
depending on the dimension of the spinor representation in the
$D$-dimensional space.\\

\noindent
The above price has to be paid anyway, on either
negatively or positively curved manifolds,  
when one 
goes beyond $D=6$.
It can be shown that $D=6$ is the maximum dimension 
possible to end up with a 4-dimensional Weyl spinor (starting from a 
6-dimensional 
Weyl), without extra degress of freedom. 
Starting from an irreducible spin 
represenation of $SO(1,D-1)$, and after some algebra,
the resulting number of four dimensional Weyl spinors is: 
$(2^{(D-5)/2}\times n)$ for $D$ odd (although 
special care should be taken in order to define spinors
in odd dimensions \cite{lawson}), 
and $(2^{D/2 -3}\times n)$
for $D$ even, where $n$ is the number of 
zero modes of the 
Dirac operator in the internal space (on a compact manifold, the 
eigenstates are all square-integrable). This number $n$ 
depends on the coupling of spinors to background fields.
Hence, the number of the
zeroeth KK fermionic modes will increase, possibly leading to a
variant number of flavours.\\

\noindent
A common way to get rid of the above extra spinorial degrees of
freedom is to use a localization mechanism as first proposed by
\cite{jackiw}. These mechanisms rely on the existence of more that 
one zero mode of the Dirac operator in the internal space, such 
that at least one of them is {\it not} normalizable. Therefore, 
a necessary
condition for applying such scenarios is to have a {\it non}-compact
internal space, because all the modes of a given Dirac operator
on a compact space are normalizable.
Hence, 
the recently discussed mechanisms 
\cite{rub,dvali,kaplan,arkani}, break down for the compact manifold, 
and an extra
care is needed for dealing with the additional modes 
(specially the zero ones). Finally, we mention that the arguments contained
here, concerning the zero KK modes, are generic in the sense that 
they do not depend on whether the product is tensor or warp.

\subsection{Massive fermions: Lower bounds}
As it is the case for the Laplacian, the eigenvalues of the Dirac
operator on a compact space are discrete. Therefore, the eigenvalues
of the squared Dirac operator are discrete and positive, and in 
addition {\it any} eigenvalue, $\nu_q^2$,is 
bounded from below by the curvature \cite{hijazi}, 
including $\nu_1^2$, 
\begin{equation}
\nu_1^2\geq \frac{d}{4(d-1)}\lambda_1
\label{nu}
\end{equation}
where $\lambda_1$ is the first eigenvalue of the Yamabe operator, 
$$
L\equiv \frac{4(d-1)}{d-2}\Delta_Y+\kappa
$$
with $\Delta_Y$ being the positive Laplacian acting on functions.
The implication of the appearance of the Laplacian 
once again in this fermionic context is
that there will be an input from the bosonic spectrum 
(as it transpires from (\ref{nu}) and 
(\ref{eq:w}) above)
in setting the bound on the massive fermionic excitations.
Therefore, the bounds on spin 1/2 and spin 0 masses are not totally 
independent   
$$
\nu_1^2\geq \frac{d}{d-2}\left[\frac{\pi^2}{4\sigma^2}-
\mbox{max} \{-(d-1)K, 0\}\right]+\kappa
$$
So, for positive curvature ($K>0$)
$$
\nu_1^2\geq \left(\frac{d}{d-2}\right)\frac{\pi^2}{4\sigma^2}+\kappa
$$
and for negative curvature ($K<0$)
$$
\nu_1^2\geq \frac{d}{d-2}\left[\frac{\pi^2}{4\sigma^2}+(d-1)K\right]+
\kappa
$$
In the the tensor product case, the above bounds read the same
for the 4-dimensional masses, $\mu_q$'s. 
Thus by choosing $\sigma^{-1}\sim{\cal O}(\mbox{TeV})$,
we find that it is natural to achieve $\mu_1^2\geq \mbox{TeV}^2$.
In the case $\kappa\geq 0$, all $\mu_q^2\gtrsim \mbox{TeV}^2$ without
any specific value of the curvature. However, when $\kappa<0$
the curvature
should satisfy an inequality similar to (\ref{kappa})
$$
\kappa\approx |(d-1)K|\lesssim \left(\frac{\pi^2d}{d-2}-1
\right)\mbox{TeV}^2
$$
It is again remarkable that $\mu_1^2\gtrsim \mbox{TeV}^2$ 
can be naturally achieved having
all our mass parameter of the same order of the compactification
mass scale.
As can be seen from the above inequalities, both $\sigma$ and 
the curvature explicitly enter the 
expressions of the bounds, and hence set the 
compactification mass scale.\\

\subsection{Massive fermions: Upper bounds}
Again, as in for $\Delta$ eigenmodes, 
an upper bound on $\nu_q^2$ exists, 
$$
\nu_q^2\leq Cq^{2/d}
$$
where $C$ is a constant that depends only on the geometry of $Y$
(even in the presence of a gauge field) \cite{vafa}. Again here
we find restrictions, though not as explicit as in 
(\ref{l1},\ref{l2},\ref{l3}), 
which limit our freedom in pushing up the KK masses arbitrarily high.\\

\noindent
All the above observations, concerning both the upper and lower bounds, 
have been done in the tensor product case though the comments on zero 
modes apply
equally to both types. 
However, if the product is warp, then the bounds and fermion 
masses will be dressed by 
the factor $e^{-\phi/2}$, as seen from (\ref{eq:d}),
and similar arguments to 
the ones carried in section (3.2) apply.

\section{Conclusions}
We considered, on general grounds, a model of Einstein gravity 
coupled to 
a Dirac
spinor and a Yang-Mills gauge theory
on $W=M_4\otimes Y$, where $Y$ is a compact internal manifold
with a scalar curvature bounded from below, and $M_4$ is our four 
dimensional world. Both tensor product and warp product are discussed.
Bounds and estimates on the masses of the effective four dimensional 
theory at the classical level  
have been pointed out. Topological restrictions in choosing
the internal manifold have been identified in order to avoid 
having certain bosonic massless modes in the four dimensional 
spectrum. In addition an upper bound on the curvature of
$Y$ has been proposed, in the case of non-positive curvature.
Geometrical {\it upper} and {\it lower} bounds have been 
presented for both bosons and fermions masses.
In the tensor product case, the characteristic
compactification mass scale 
for bosons is the diameter of the internal 
manifold, $\sigma^{-1}$, along with $|K|$ when the curvature 
$\kappa<0$. For fermions, the compactification mass scale is always
set by $\sigma$ and the curvature, and this is due
to an input from the bosonic spectrum 
in setting the bound on the massive fermionic excitations.
Therefore, there is an interplay between spin 1/2 and spin 0 
sectors.\\

\noindent
For both fermions and bosons, it turns out that having the masses
of the lowest excitations $\gtrsim \mbox{TeV}$ is naturally achieved 
by taking all the dimensionful parameters, arising from compactification,
to be $\sim {\cal O}(\mbox{TeV})$ (no fine-tuning required). 
In the warp product case, no direct bounds can be applied for massive
states without 
the knowledge of both the specific shape of the warp factor and 
the field dependence on the internal space, though we are able to 
implement general estimates.
``Zero-mode'' 
arguments can be applied to both kinds of products.
From the analysis conducted in this work, we conclude that 
non-positively curved internal manifolds with $b_0=1$ and $b_1=0$
are strongly favoured for phenomenological purposes.\\

\noindent
Finally, a comment about {\it non-compact} internal manifolds: 
it has been argued
\cite{gz} that the spectrum of the 
Laplacian on non-compact spaces
of finite volume has a discrete sector. Moreover, it has 
been shown recently \cite{b7} that suitable choice of spin
structure also leads to a discrete spectrum of the Dirac operator 
for {\it non-compact} hyperbolic manifolds of finite volume.
One can therefore contemplate analyzing similar bounds
for such non-compact spaces, along the same lines as we have executed 
in this work, and discuss their phenomenological implications \cite{a}.\\

{\it Acknowledgement.}
One of us (R.T.) would like to thank Seif Randjbar-Daemi
and George Thompson for useful discussions.

\subsection*{Appendix}
The expressions of the Laplacian acting on various tensors 
have been worked out in
\cite{lich2}, and for convenience we list 
here the relevant expressions. 
\begin{eqnarray}
\Delta \alpha&=& -\nabla_i\nabla^i \alpha=-\frac{1}{\sqrt{g}}\partial
(\sqrt{g} g^{ik}\partial_k)\alpha \nonumber\\
(\Delta \alpha)_r&=&-\nabla_i\nabla^i \alpha_r
-{\cal R}^h_r  \alpha_h
\label{lapl}
\\
(\Delta \alpha)_{kl}&=&-\nabla_i\nabla^i \alpha_{kl}
+{\cal R}^h_k  \alpha_{hl} +{\cal R}^h_l \alpha_{kh} -2
{\cal R}_{ki,lh} \alpha^{ih}\nonumber
\end{eqnarray}
where $i,j,...etc=1,...,$dimension of the manifold on which the tensors
and Laplacians are 
defined. ${\cal R}^h_k$ and ${\cal R}_{ki,lh}$
are Ricci and Riemann tensors respectively. 


\end{document}